\documentstyle[aps,preprint]{revtex}
\begin{document}
\preprint{SNUTP-96115 ~~gr-qc/9611040}

\title{Problem of Unitarity and Quantum Corrections
in Semiclassical Quantum Gravity}

\author{Sang Pyo Kim}
\address{Department of Physics\\
Kunsan National University \\
Kunsan 573-701, Korea}
\maketitle
\begin{abstract}
Using both the Born-Oppenheimer idea and the
de Broglie-Bohm interpretation of wavefunction
we represent in a different way the semiclassical quantum gravity
from the Wheeler-DeWitt equation in an oscillating regime
which can preserve completely the unitary quantum
evolution of a matter field at the expense of a
nonlinear gravitational field equation, but
has the same asymptotic limit as the others.
We apply the de Broglie-Bohm interpretation to the nonlinear
gravitational field equation
to develop a perturbation method to find the quantum corrections
of a matter field to the gravity.
The semiclassical Einstein equation with the quantum corrections
is found for a minimal quantum FRW cosmological model.
\end{abstract}

\section{Introduction}

The problem of unitarity and back-reaction of a matter
field in a curved spacetime
has been an important and critical issue of gravity
such as black holes and the very early Universe.
Although the complete solution of this problem should be found
within the context of the right and consistent quantum gravity,
there has been quite recently an attempt
to tackle this problem from the canonical quantum gravity based on the
Wheeler-DeWitt (WDW) equation \cite{kiefer}.
In the canonical quantum gravity approach to  the gravity
plus matter system, the gravitational wavefunctions of the WDW equation
are peaked along (semi)classical trajectories of
the (semi)classical Einstein equation, and the matter
field, which obeys a time-dependent functional Schr\"{o}dinger equation,
evolves quantum mechanically along each trajectory.
These two equations constitute the so-called semiclassical quantum gravity.
We shall distinguish this approach from the conventional
field theoretic approach in the curved spacetime \cite{birrel}.
Recently the semiclassical quantum gravity has been applied
to a two dimensional black hole model \cite{alwis}.

One can derive the semiclassical quantum gravity in many different ways,
for instance, depending on whether or not he or she includes the back-reaction
of the matter field to the semiclassical Einstein equation in the
lowest order of the power series expansion in terms of the inverse Planck
mass. When one searches for a WKB-type wavefunction for the
entire WDW equation and expands its total action in the inverse power series of
the Planck mass, he or she obtains the vacuum (classical) Einstein equation
and the time-dependent functional Schr\"{o}dinger equation
for the matter field on it \cite{banks}.
On the other hand, when one adopts the Born-Oppenheimer idea
by treating the gravitational field as a massive particle
degree of freedom and the matter field  as a light particle degree of freedom,
he or she can separate two different mass scales
and derive the semiclassical Einstein equation with the quantum
expectation value of the energy-momentum tensor operator as a source of matter
and the time-dependent Schr\"{o}dinger equation
for the matter field \cite{brout,kk}.
In previous papers \cite{kim1}, by separating and expanding the WDW equation
in the inverse power series of the Planck mass,
we were able to derive the semiclassical quantum gravity
and to find the quantum states of the matter field
in the asymptotic limit of large Planck mass.
Quite recently Bertoni {\it et al} \cite{bertoni} have shown
that the semiclassical quantum gravity
represented in three different ways are in fact equivalent  to
each other in the asymptotic limit ${\cal O}
\Bigl(\frac{\hbar}{m_P} \Bigr)$.

In the asymptotic limit of large Planck mass,
the semiclassical quantum gravity is unitary
in the sense of conservation of probability of the matter wavefunctions
\cite{kim1}.
In a different representation, there are, however, unitarity violating terms
in the time-dependent Schr\"{o}dinger equation
for the matter field beyond the asymptotic limit \cite{kiefer2}.
These unitarity violating terms come from the slowly varying
amplitude of the gravitational wavefunction in the expansion of the
WDW equation in the inverse power series of the Planck mass.
The origin of these unitarity violating terms
may not, however, follow inherently from the WDW equation.
Bertoni {\it et al} \cite{bertoni} have further shown that the semiclassical
quantum gravity can indeed preserve the unitarity of quantum fields
provided that one can define the cosmological time defined
through the Hamilton-Jacobi equation of gravitational field
including the back-reaction of matter field.
A remark on the different representations of almost the same semiclassical
quantum gravity is that the semiclassical quantum gravity in Ref. \cite{kim1}
is based on the complicated matrix equation of the WDW
equation which incorporates actually the scattering of Cauchy data
at the second or third quantized level and also considers linear combinations
of the wave functions \cite{kim2},
in contrast with one wave function in Refs. \cite{brout} and \cite{bertoni}.
So, it would be worthy to show the unitarity of quantum fields
from the WDW equation using the equivalent matrix equation \cite{kk,kim1}.

In this paper we observe that the WDW equation is not uniquely separated
into the gravitational field equation and the time-dependent
Schr\"{o}dinger equation according to the different mass scales following
only the Born-Oppenheimer idea \cite{kim1}, and that, especially when
we apply de Broglie-Bohm interpretation \cite{holland} at the same time
to the wavefunction of the gravitational field equation, this
arbitrariness in  the separation of the WDW equation in an oscillating regime
can also allow the semiclassical quantum gravity which consists
of a nonlinear gravitational field equation and a time-dependent
Schr\"{o}dinger equation that
indeed does {\it preserve} the norm of the quantum states of matter field
during the quantum evolution. As in Ref. \cite{bertoni},
we assume no asymptotic limit to derive the semiclassical quantum gravity.
However, in the asymptotic limit, it reduces identically to the
semiclassical quantum gravity in previous papers \cite{brout,kim1}.
Thus, the question of unitarity of the quantum evolution of matter fields
may not be settled within the canonical quantum gravity, and should be answered
by  more fundamental theory.
Furthermore, we shall use this semiclassical quantum gravity
to find the higher order quantum corrections of
the matter field to the gravity.
Finally, we shall apply the semiclassical quantum
gravity to the quantum Friedmann-Robertson-Walker
(FRW) cosmological model coupled to a minimal scalar field.

The organization of the paper is as follows:
In Sect. II we derive the nonlinear gravitational
field equation and the time-dependent Schr\"{o}dinger equation,
and show that they preserve the unitarity of the
quantum evolution of matter field.
In Sect. III we develop a perturbation method for
the semiclassical Einstein  equation with the quantum potential and
the higher order quantum corrections of matter field.
In Sect. IV we apply the semiclassical
quantum gravity to the quantum FRW cosmological model to find
the effective energy density.

\section{Nonlinear Gravitational Field Equation
and Unitarity}

We shall follow the Born-Oppenheimer idea to separate the WDW equation
into the nonlinear gravitational field equation and the
time-dependent Schr\"{o}dinger equation. The canonical quantum gravity
for a gravity coupled to a matter field, typically represented
by a scalar field, is described by the WDW equation
\begin{equation}
\left[- \frac{\hbar^2}{2m_P}
\nabla^2 - m_P V(h_a) +
\hat{\bf H} \left(\frac{i}{\hbar}
\frac{\delta}{\delta \phi}, \phi, h_a \right)
\right] \Psi (h_a, \phi) = 0,
\label{wd eq}
\end{equation}
where $h_a$ represents the gravitational field
and $\phi$ the matter field. Here $\nabla^2$ is
the Laplace-Beltrami operator in superspace with
the signature $(-, +, \cdots, +)$.
Take the Born-Oppenheimer-type
wavefunction for the WDW equation
\begin{equation}
\left| \Psi(h_a, \phi)\right>
= \psi(h_a) \left| \Phi(\phi, h_a) \right>,
\label{wa fuc}
\end{equation}
where $\psi$ depends only on the gravitational field
and $\Phi$ both on  the matter field and
on the gravitational field as an explicit parameter.
The Born-Oppenheimer idea assumes that the quantum states of matter
field belong to a Hilbert space that varies continuously
on the superspace of the gravitational field.
The validity of the form of wavefunction (\ref{wa fuc})
should be checked by the unitarity of the quantum state
of matter field in the semiclassical quantum gravity.

Before we delve into representing in a different way
the semiclassical quantum gravity that preserves the unitarity
of the quantum matter field,
we point out a major similarity and a minor difference among the
approaches to the semiclassical quantum gravity
in Refs. \cite{kk,kim1} and \cite{bertoni}.
The similarity is that, as explained in the Introduction, most of
the approaches \cite{brout,kk,kim1,bertoni} take into account
the expectation value of energy-momentum tensor as its matter source
in the semiclassical Einstein equation. This is the quintessence of the
Born-Oppenheimer idea originally applied to a molecular system,
but in Ref. \cite{banks} the back-reaction was excluded at the
lowest order by expanding the total action in the inverse power
series of the Planck mass.

The difference comes from the interpretation of the WDW equation (\ref{wd eq})
whether as the zero-energy Schr\"{o}dinger equation for
a coupled system \cite{brout,bertoni} or as
a functional wave equation describing the evolution (scattering) of Cauchy data
from an initial spacelike hypersurface, usually assumed to be the early stage
of the Universe, to a later spacelike hypersurface \cite{kim2}.
In a toy model without a matter field, the WDW equation may apparently look
like the Schr\"{o}dinger equation with zero energy,
but with additional inhomogeneous gravitational degrees
of freedom or matter field degrees of freedom, Eq. (\ref{wd eq})
has the Minkowski signature $(-. +, \cdots, +)$.
The exact wave functions for the quantum FRW cosmological model
minimally coupled to a scalar field with a general potential, which
is to be considered in Sec. IV,  are extremely difficult to find due
to the non-separability of quantum states of the
scalar field from the gravitational field and have not been found yet.
In a systematic way we can make use of the  initial value problem for
Eq. (\ref{wd eq}) formulated in a matrix form to describe
the evolution (scattering) of Cauchy data
from one spacelike hypersurface to another, including the evolution of
any quantum state. The matrix gravitational field equations
in Ref. \cite{kk} indeed incorporated the Cauchy data.
The semiclassical quantum gravity in previous papers \cite{kim1}
was based on this exact matrix formulation for the canonical quantum
gravity, and took into account not only single wave function but also
a linear combination of wave functions.

Following Refs. \cite{kk,kim1}, we may expand any quantum state
with respect to a suitably chosen orthonormal basis relevant
to the matter field:
\begin{equation}
\left| \Phi(\phi, h_a) \right> = \sum_{k} c_k (h_a)
\left| \Phi_k (\phi, h_a) \right>.
\label{quan st}
\end{equation}
Substitute Eqs. (\ref{wa fuc}) and (\ref{quan st})
into the WDW equation (\ref{wd eq})
to get the matrix equation \cite{kim1}
\begin{eqnarray}
&&c_n (h_a) \Bigl( - \frac{\hbar^2}{2m_P}
\nabla^2 - m_P V(h_a) + {\bf H}_{nn} (h_a) \Bigr)
\psi(h_a) \nonumber\\
&& ~~~~~~+ \Biggl(
- \frac{\hbar^2}{m_P} \nabla \psi (h_a)
\cdot \Bigl( \nabla c_n (h_a)
- i \sum_{k} {\bf A}_{nk} (h_a) c_k (h_a)
\Bigr) \nonumber\\
&& ~~~~~~~~~~~~+ \psi (h_a) \Bigl(
\sum_{k \neq n} {\bf H}_{nk}(h_a) c_k (h_a)
- \frac{\hbar^2}{2m_P} \sum_{k}
{\bf \Omega}_{nk} (h_a) c_k (h_a)
\Bigr)
\Biggr) = 0.
\label{mat eq}
\end{eqnarray}
where
\begin{eqnarray}
{\bf H}_{nk} (h_a) :&=& \left<\Phi_n\right|
\hat{\bf H} \left|\Phi_k \right>,\nonumber\\
{\bf A}_{nk} (h_a) :&=& i \left<\Phi_n\right|
\nabla \left|\Phi_k \right>,\nonumber\\
{\bf \Omega}_{nk} (h_a) :&=&
\nabla^2 \delta_{nk} - 2i {\bf A}_{nk}
\cdot \nabla +
\left<\Phi_n\right|
\nabla^2 \left|\Phi_k \right>.
\end{eqnarray}
This is an intermediate step for
the derivation of the semiclassical quantum gravity.

We now derive the semiclassical quantum gravity
in a different way from the previous papers \cite{kim1},
so that it can preserve
the unitarity of quantum states completely.
A complex solution of the gravitational field  equation
in an oscillating regime
can always be rewritten in the form
\begin{equation}
\psi(h_a) = \left( \psi^* \psi \right)^{\frac{1}{2}}
\exp \left( \frac{i}{\hbar} S(h_a) \right).
\label{com sol}
\end{equation}
Substituting the complex solution (\ref{com sol})
into the matrix equation (\ref{mat eq}) and
assuming $c_n \neq 0$,
we obtain the two coupled equations:
\begin{eqnarray}
 \Biggl( - \frac{\hbar^2}{2m_P}
\nabla^2 &-& m_P V(h_a) + {\bf H}_{nn} (h_a)
\nonumber\\
&-& \frac{\hbar^2}{m_P}
\frac{\nabla (\psi^* \psi)^{\frac{1}{2}}}{(\psi^* \psi)^{\frac{1}{2}}}
\cdot \Bigl(\frac{\nabla c_n (h_a)}{c_n (h_a)}
-  i \sum_{k}  {\bf A}_{nk} (h_a)
\frac{c_k (h_a)}{c_n (h_a)}  \Bigr) \Biggr) \psi (h_a) = 0
\label{grav eq}
\end{eqnarray}
and
\begin{eqnarray}
i \frac{\hbar}{m_P} \nabla S(h_a)
\cdot \nabla c_n (h_a)
&+& \frac{\hbar}{m_P} \nabla S(h_a)
\cdot \sum_{k} {\bf A}_{nk} (h_a) c_k (h_a)
\nonumber\\
&-& \sum_{k \neq n} {\bf H}_{nk}(h_a) c_k (h_a)
+ \frac{\hbar^2}{2 m_P} \sum_{k}
{\bf \Omega}_{nk} (h_a) c_k (h_a) = 0.
\label{matt eq}
\end{eqnarray}
Eq. (\ref{grav eq}) may be interpreted as
the nonlinear gravitational field
equation with the quantum
back-reaction of the matter field.
In terms of the cosmological time
\begin{equation}
\frac{\delta}{\delta \tau} :=
\frac{1}{m_P} \nabla S(h_a) \cdot \nabla ,
\end{equation}
Eq. (\ref{matt eq}) can be rewritten as
\begin{eqnarray}
i \hbar \frac{\delta}{\delta \tau} c_n (h_a)
= &&\sum_{k \neq n} {\bf H}_{nk}(h_a) c_k (h_a)
\nonumber\\
&-&  \frac{\hbar}{m_P} \nabla S(h_a)
\cdot \sum_{k} {\bf A}_{nk} (h_a) c_k (h_a)
- \frac{\hbar^2}{2 m_P} \sum_{k}
{\bf \Omega}_{nk} (h_a) c_k (h_a).
\label{ts eq}
\end{eqnarray}
Eq. (\ref{ts eq}) may now be interpreted as
the time-dependent functional Schr\"{o}dinger
equation for  the matter field on the spacetime
determined by Eq. (\ref{grav eq}).
It should be noted that there can be some arbitrariness
in separating the WDW equation.
Instead of ${\bf H}_{nn}$ in Eq. (\ref{grav eq})
one may use a quantum back-reaction
${\bf K}_{nn} (h_a)$. Then, the nonlinear
gravitational field equation (\ref{grav eq})
is replaced with ${\bf K}_{nn}$.
We rewrite the matter field equation
by a vector notation
\begin{equation}
i \hbar \frac{\delta}{\delta \tau} {\bf c} (h_a (\tau))
= {\bf M} (h_a (\tau)) {\bf c} (h_a)(\tau))
\end{equation}
where ${\bf M}$ is the hermitian matrix
\begin{equation}
{\bf M} (h_a (\tau)) = {\bf M}_0 (h_a (\tau)) + {\bf M}_1 (h_a (\tau)),
\end{equation}
consisted of
\begin{eqnarray}
{\bf M}_0 (h_a (\tau)) &=& \sum_{k \neq n} {\bf H}_{nk}(h_a(\tau))
-  \frac{\hbar}{m_P} \nabla S(h_a(\tau))
\cdot \sum_{k} {\bf A}_{nk} (h_a(\tau))
- \frac{\hbar^2}{2 m_P} \sum_{k}
{\bf \Omega}_{nk} (h_a (\tau)),
\nonumber\\
{\bf M}_1 (h_a (\tau)) &=& {\bf K}_{nn} (h_a(\tau))
- {\bf H}_{nn} (h_a(\tau)).
\end{eqnarray}
The unitarity of the quantum matter
field is always fully respected
\begin{equation}
{\bf c}^\dagger (\tau) \cdot {\bf c} (\tau) = 1.
\end{equation}

A minor difference of this paper from other related
works \cite{banks,brout,kiefer2} is that we followed
closely Ref. \cite{kim2}, which interpreted the WDW equation
as the quantum scattering of the Cauchy data including the quantum states
of matter fields. The difference from previous works \cite{kim1}
lies on the fact that we made use of the arbitrariness
in order to separate the nonlinear gravitational field equation
(\ref{grav eq}) from the WDW equation and the time-dependent
Schr\"{o}dinger equation (\ref{ts eq}) that
preserves the unitarity of the quantum evolution of matter field.
The unitarity of quantum field in the oscillating regime
has been first proved by Bertoni {\it et al} \cite{bertoni}
using the semiclassical quantum gravity in Ref. \cite{brout}.

\section{Quantum Corrections}

We apply the de Broglie-Bohm interpretation of wavefunctions
to the nonlinear gravitational field equation (\ref{grav eq}).
The main idea of the de Broglie-Bohm interpretation is that a time-dependent
Schr\"{o}dinger equation is mathematically equivalent to a pair of equations,
the time-dependent Hamilton-Jacobi equation with an additional
quantum potential and the equation for the conservation
of probability \cite{holland}.
The concept of trajectories with the quantum potential
provides us with all the same physical predictions as
the standard quantum mechanics.
Thus, the de Broglie-Bohm interpretation is another way of interpreting
quantum mechanics, not a kind of approximation methods
such as the WKB method, even though the de Broglie-Bohm interpretation
has the WKB as an asymptotic limit, when the quantum potential is negligible.

According to the de Broglie-Bohm interpretation,
we separate the real and the imaginary
parts of the nonlinear gravitational field equation.
Thus, by substituting  $f \equiv (\psi^* \psi)^{1/2}$
and equating the real and the imaginary parts  of Eq. (\ref{grav eq}),
we obtain
\begin{eqnarray}
\frac{1}{2m_P} \left(\nabla S \right)^2 - m_P V
+ {\bf H}_{nn}
 - \frac{\hbar^2}{2 m_P} \frac{\nabla^2 F}{F}
 - \frac{\hbar^2}{m_P} {\rm Re}
 \bigl({\bf Q}_{nn} \bigr) = 0,
\label{hj eq}
\\
\frac{1}{2} \nabla^2 S +
\frac{\nabla F}{F} \cdot
\nabla S
+ {\rm Im} \big({\bf Q}_{nn} \bigr) = 0,
\label{prob eq}
\end{eqnarray}
where
\begin{equation}
{\bf Q}_{nn} =
\frac{\nabla F}{F} \cdot
 \Bigl( \frac{\nabla c_n}{c_n}
 - i \sum_{k} {\bf A}_{nk} \frac{c_k}{c_n} \Bigr).
 \end{equation}
If there is no parametric-coupling of gravity with matter,
the following term
\begin{equation}
V_{quant} :=
- \frac{\hbar^2}{2 m_P} \frac{\nabla^2 F}{F}
\end{equation}
is the quantum potential, and Eq. (\ref{prob eq}) is
the equation for the probability conservation \cite{holland}.
The last terms in Eqs. (\ref{hj eq}) and (\ref{prob eq})
are the quantum back-reaction of matter field to the gravity.

In the asymptotic limit ${\cal O} \Bigl(\frac{\hbar}{m_P} \Bigr)$,
the matter field equation becomes
\begin{equation}
i \hbar \frac{\delta}{\delta \tau} c_{n}^{(0)} (h_a)
= \sum_{k \neq n} {\bf H}_{nk}(h_a) c_{k}^{(0)} (h_a)
-  \frac{\hbar}{m_P} \nabla S^{(0)}(h_a)
\cdot \sum_{k} {\bf A}_{nk} (h_a) c_{k}^{(0)} (h_a)
\end{equation}
and the gravitational field equation reduces
to the Hamilton-Jacobi equation of the zeroth order
\begin{equation}
\frac{1}{m_P} \left(\nabla S^{(0)} \right)^2 - m_P V
+ {\bf H}_{nn} = 0.
\end{equation}
As an orthonormal basis, we choose the exact quantum states
of the time-dependent functional Schr\"{o}dinger equation
\begin{equation}
i \hbar \frac{\delta}{\delta \tau}
\bigl|\Phi^{(0)} (\phi, \tau) \bigr> =
\hat{\bf H} \Bigl(\frac{i}{\hbar}
\frac{\delta}{\delta \phi}, \phi, h_a (\tau)
\Bigr) \bigl|\Phi^{(0)} (\phi, \tau) \bigr>.
\end{equation}
This orthonormal basis simplifies considerably
the algebra; especially, the matter field equation becomes
\begin{equation}
i \hbar \frac{\delta}{\delta \tau} c_{n}^{(0)} (h_a)
= -  \frac{\hbar}{m_P} \nabla S^{(0)}(h_a)
\cdot {\bf A}_{nn} (h_a) c_{n}^{(0)} (h_a).
\end{equation}
The solution is
\begin{equation}
c_n^{(0)} = \exp \Bigl(i \int ({\bf A}_{nn})_a     dh_a \Bigr).
\end{equation}

To find the gravitational field wavefunction and the quantum states
of matter field, we use a perturbation method
that expands the coefficient functions and the gravitational action
as
\begin{eqnarray}
c_n (h_a) &=& \sum_{p = 0}^{\infty}
\Bigl(\frac{\hbar}{m_P} \Bigr)^p c_n^{(p)} (h_a),
\nonumber\\
c_k (h_a) &=& \sum_{p = 1}^{\infty}
\Bigl(\frac{\hbar}{m_P} \Bigr)^p c_k^{(p)} (h_a),
{}~(k \neq n),
\end{eqnarray}
and
\begin{eqnarray}
S (h_a) &=& \sum_{p = 0}^{\infty}
\Bigl(\frac{\hbar}{m_P} \Bigr)^p S^{(p)} (h_a),
\nonumber\\
F (h_a) &=& \sum_{p = 0}^{\infty}
\Bigl(\frac{\hbar}{m_P} \Bigr)^p f^{(p)} (h_a).
\end{eqnarray}
The dominant contribution to the quantum state
comes from $c_n$. The other terms $c_k, (k\neq n)$
are of the order of ${\cal O} \Bigl(\frac{\hbar}{m_P} \Bigr)$.
The prefactor $f$ of the gravitational wavefunction is
determined by
\begin{equation}
\frac{\nabla f^{(0)}}{f^{(0)}} \cdot
\nabla S^{(0)}
= - \frac{1}{2} \nabla^2 S^{(0)},
\end{equation}
whose solution is
\begin{equation}
f^{(0)} = \frac{1}{ \bigl(\nabla S^{(0)} \cdot
\nabla S^{(0)} \bigr)^{1/4}}.
\end{equation}
The first order coefficient functions
of the quantum state are
\begin{equation}
i  \frac{\delta}{\delta \tau} c_{k}^{(1)} (h_a)
= - \frac{1}{2} {\bf \Omega}_{kn} (h_a) c_{n}^{(0)} (h_a),
\end{equation}
where
\begin{equation}
{\bf \Omega}_{kn} (h_a) c_{n}^{(0)} (h_a)
= \Bigl(2 {\bf A}_{nk} \cdot {\bf A}_{nn}
- {\bf A}_{nn} \cdot {\bf A}_{kn} -
\bigl({\bf A} \cdot {\bf A} \bigr)_{nn} \Bigr)
c_n^{(0)} (h_a).
\end{equation}
We determine $S^{(1)}$  from
\begin{equation}
\frac{1}{m_P} \nabla S^{(1)} \cdot \nabla S^{(0)}
= \frac{\hbar}{2} \frac{\nabla^2 f^{(0)}}{f^{(0)}}.
\end{equation}
The procedure can be repeated to yield the higher order
quantum corrections of the matter field
to the semiclassical Einstein equation
and the exact quantum state.

\section{Quantum FRW Cosmological Model}

We now apply the semiclassical quantum gravity developed
in this paper to the frequently employed quantum FRW cosmological model.
The higher order quantum
corrections in Sect. III to the asymptotic semiclassical
quantum gravity have not yet been found explicitly
even for the FRW model,
although higher order quantum corrections were found
in the semiclassical quantum gravity whose
lowest order equation is the
vacuum Einstein equation and which does violate the unitarity
at the first order \cite{kiefer2}.
Moreover, since the quantum field preserves the unitarity
throughout the evolution, the application to the
FRW model is expected to be particularly useful in searching
for the quantum effects of matter field to the gravity.

\subsection{Minimal Scalar Field}

The simplest but nontrivial FRW quantum cosmological model
is with a minimal scalar field (inflaton).
The WDW equation is
\begin{equation}
\Biggl[
 \frac{2 \pi \hbar^2}{3 m_P a}
\frac{\partial^2}{\partial a^2}
- \frac{3 m_P}{8 \pi} V(a)
+ \hat{\bf H} \Bigl(\frac{i}{\hbar}
\frac{\delta}{\delta \phi}, \phi, a \Bigr)
\Biggr] \Psi(a, \phi) = 0,
\label{frw mod}
\end{equation}
where $a$ is the size of the universe, and
\begin{equation}
V(a) = k a - \Lambda a^3
\end{equation}
is the gravitational potential consisting of the three-curvature
and the cosmological constant,  and $\phi$ denotes the massive scalar field.
The extended supermetric is
\begin{equation}
ds^2 = -a da^2 + a^3 d\phi^2,
\end{equation}
and the rescaling $a = \Bigl(\frac{3}{4\pi} \Bigr)^{1/3}
 \tilde{a}$ recovers the WDW equation of the form in Eq. (\ref{wd eq}).
Remembering the superspace signature $(-)$,
the cosmological time is related to the gravitational action by
\begin{equation}
\frac{\partial}{\partial \tau}  =  - \frac{4 \pi}{3m_P a}
\frac{\partial S (a)}{\partial a}
\frac{\partial}{\partial a}.
\label{cos time}
\end{equation}
Then, we find that
\begin{equation}
\dot{a} (\tau) = - \frac{4 \pi}{3 m_P a} S'
\label{rel}
\end{equation}
where and hereafter overdots and primes will denote
derivative with respect to $\tau$ and $a$, respectively.
We rewrite the nonlinear gravitational field equation
as
\begin{equation}
\Biggl[ \frac{2 \pi \hbar^2}{3 m_P a}
\frac{\partial^2}{\partial a^2}
- \frac{3 m_P}{8 \pi} V(a)
+ {\bf H}_{nn}
- \frac{4 \pi \hbar^2}{3 m_P a}
\frac{F'}{F} \Bigl(\frac{c_n'}{c_n}
+ i\frac{3 m_P}{4\pi} \frac{a}{S'} \sum_{k} {\bf B}_{nk}
\frac{c_k}{c_n} \Bigr)
\Biggr] \psi(a) = 0.
\end{equation}
In terms of the cosmological time, the time-dependent functional
Schr\"{o}dinger equation is rewritten as
\begin{eqnarray}
i \hbar \frac{\partial}{\partial \tau} c_n (\tau)
= &&\sum_{k \neq n} {\bf H}_{nk}(\tau) c_k (\tau)
- \hbar  \sum_{k} {\bf B}_{nk} (\tau) c_k (\tau)
\nonumber\\
&-& \frac{2 \pi \hbar^2}{3 m_P a}  \sum_{k}
{\bf \Omega}_{nk} (\tau) c_k (\tau),
\label{frw sc}
\end{eqnarray}
where
\begin{eqnarray}
{\bf H}_{nk} (\tau) &=& \bigl< \Phi_n \bigr|
\hat{\bf H} \bigl| \Phi_k \bigr>,
\nonumber\\
{\bf B}_{nk} (\tau) &=& i \bigl< \Phi_n \bigr|
\frac{\partial}{\partial \tau}
\bigl| \Phi_k \bigr>,
\nonumber\\
{\bf \Omega}_{nk} (\tau) &=& -
\frac{1}{\dot{a}^2}
\Biggl[
\Bigl( \frac{\partial^2}{\partial \tau^2}
- \frac{\ddot{a}}{\dot{a}} \frac{\partial}{\partial \tau}
 \Bigr) \delta_{nk} - 2i {\bf B}_{nk}
 \frac{\partial}{\partial \tau}
 +
\bigl<\Phi_n \bigr|
\frac{\partial^2}{\partial \tau^2}
- \frac{\ddot{a}}{\dot{a}} \frac{\partial}{\partial \tau}
\bigl| \Phi_k \bigr> \Biggr].
\end{eqnarray}
In the above equations we made use of the relations
(\ref{cos time}) and (\ref{rel}).
Following the de Broglie-Bohm interpretation,
we substitute the wavefunction $\psi = F e^{\frac{i}{\hbar}
S}$ and separate the real and the imaginary parts of the
gravitational field equation to obtain
\begin{equation}
\frac{2 \pi}{3 m_P a}  S'^2
+ \frac{3 m_P}{8 \pi} V(a) = {\bf H}_{nn}
- \frac{4 \pi \hbar^2}{ 3m_P a \dot{a}} \frac{F'}{F}
{\rm Re} \bigl( {\bf R}_{nn} \bigr)
+ \frac{2 \pi \hbar^2}{3 m_P a}
\frac{F''}{F},
\label{frw}
\end{equation}
and
\begin{equation}
\frac{F'}{F} S' + \frac{1}{2} S'' =
\frac{\hbar}{\dot{a}} \frac{F'}{F} {\rm Im}
\bigl( {\bf R}_{nn} \bigr),
\label{pre fac}
\end{equation}
where
\begin{equation}
{\bf R}_{nn} = \frac{\dot{c}_n }{c_n} - i \sum_k
{\bf B}_{nk} \frac{c_k}{c_n},
\end{equation}
and ${\bf Q}_{nn} = \frac{F'}{F} {\bf R}_{nn}$.
We solve Eq. (\ref{pre fac}), and let
\begin{equation}
{\bf U}_{nn} := \frac{F'}{F} =
- \frac{1}{2}
\frac{ \bigl( a \dot{a} \bigr)^{\cdot} }{
a \dot{a}^2 + \frac{ 4 \pi \hbar}{3 m_P} {\rm Im}
\bigl( {\bf R}_{nn} \bigr) },
\label{f/f}
\end{equation}
where we again used (\ref{cos time}) and (\ref{rel}).
Using again (\ref{rel}) and (\ref{f/f}),
we rewrite Eq. (\ref{frw}) as
\begin{equation}
\Bigl(\frac{\dot{a}}{a} \Bigr)^2
+ \frac{1}{a^3} V(a) = \frac{8 \pi}{3 m_P a^3} \Biggl[
{\bf H}_{nn}
- \frac{4 \pi \hbar^2}{ 3m_P a \dot{a}} {\bf U}_{nn} {\rm Re}
\bigl( {\bf R}_{nn} \bigr) +
\frac{2 \pi \hbar^2}{3 m_P a}  \Bigl({\bf U}^2_{nn}
+ \frac{1}{\dot{a}} \dot{\bf U}_{nn} \Bigr) \Biggr].
\label{frw 1}
\end{equation}
We can interpret Eq. (\ref{frw 1}) as the semiclassical Einstein
equation with the quantum back-reaction of the matter field and
the higher order quantum corrections from the fluctuations
of matter field and geometry itself.
In this sense,
\begin{equation}
{\bf T}_{nn} :=
{\bf H}_{nn}
- \frac{4 \pi \hbar^2}{ 3m_P a \dot{a}}
 {\bf U}_{nn} {\rm Re}
\bigl( {\bf R}_{nn} \bigr) +
\frac{2 \pi \hbar^2}{3 m_P a}  \Bigl( {\bf U}^2_{nn}
+ \frac{1}{\dot{a}} \dot{\bf U}_{nn} \Bigr)
\end{equation}
is the effective energy density from the quantum fluctuation of matter
fields.
Eq. (\ref{frw sc}) is the time-dependent functional Schr\"{o}dinger equation
for the matter field on the spacetime determined by Eq. (\ref{frw 1}).

\subsection{Massive Scalar Field}

We work out explicitly the specific case of a
massive scalar field.
The Hamiltonian of the massive scalar field is
\begin{equation}
 \hat{\bf H} \Bigl(\frac{i}{\hbar}
\frac{\delta}{\delta \phi}, \phi, a \Bigr)
=
- \frac{\hbar^2}{2 a^3} \frac{\partial^2}{\partial \phi^2}
+ \frac{a^3 m^2}{2} \phi^2.
\end{equation}
In Ref. \cite{kim3}
a Fock space-representation of quantum states
satisfying the time-dependent functional Schr\"{o}dinger equation
was constructed in terms of the annihilation and creation operators
\begin{equation}
\hat{b} (\tau) =  \phi_c^* (\tau) \hat{\pi}_{\phi}
- a^3 (\tau) \dot{\phi}_c^* (\tau) \hat{\phi},~
\hat{b}^{\dagger} (\tau) = \bigl( \hat{b} (\tau) \bigr)^{\dagger}
\end{equation}
where $\phi_c$ is a complex solution of the
classical field equation
\begin{equation}
\ddot{\phi}_c (\tau) + 3 \frac{\dot{a} (\tau)}{a(\tau)}
\dot{\phi}_c (\tau) + m^2 \phi_c (\tau) = 0.
\end{equation}
The effective gauge potential is found in an operator form:
\begin{equation}
{\bf B} = \alpha (\tau) \hat{b}^{\dagger} (\tau)
\hat{b} (\tau)
+ \beta (\tau) \hat{b}^2 (\tau) +
\beta^* (\tau) \hat{b}^{\dagger 2} (\tau),
\end{equation}
where
\begin{eqnarray}
\alpha (\tau) &=& \hbar a^3 (\tau) \bigl(
\dot{\phi}^* (\tau) \dot{\phi} (\tau)
+ m^2 \phi^* (\tau) \phi (\tau) \bigr)
\nonumber\\
\beta (\tau) &=& - \frac{\hbar a^3}{2}
\bigl(\dot{\phi}^2 (\tau) + m^2 \phi^2 (\tau) \bigr)
\end{eqnarray}
Since $c_n$ is of the order ${\cal O} (1)$
but $c_k (k \neq n) $  of the order
${\cal O} \Bigl(\frac{\hbar}{m_P} \Bigr)$,
we obtain the coefficient function
\begin{equation}
c_n^{(0)} (\tau) = \exp \Bigl(i \int {\bf B}_{nn}
(\tau) d\tau \Bigr).
\end{equation}
The lowest contributions to ${\bf R}_{nn}$ and ${\bf U}_{nn}$
are found
\begin{eqnarray}
{\bf R}_{nn}^{(0)} &=& 0,
\nonumber\\
{\bf U}_{nn}^{(0)} &=&
- \frac{1}{2}
\frac{ \bigl( a \dot{a} \bigr)^{\cdot} }{
a \dot{a}^2}.
\end{eqnarray}
Thus, the semiclassical Einstein equation at this order becomes
\begin{equation}
\Bigl(\frac{\dot{a}}{a} \Bigr)^2
+ \frac{1}{a^3} V(a) = \frac{8 \pi}{3 m_P a^3} \Biggl[
{\bf H}_{nn} +
\frac{2 \pi \hbar^2}{3 m_P a}
\Bigl( {\bf U}^{(0)2}_{nn}
+ \frac{1}{\dot{a}} \dot{\bf U}^{(0)}_{nn} \Bigr) \Biggr],
\end{equation}
where
\begin{equation}
{\bf H}_{nn} = \hbar a^3 (\tau)
\Bigl( n + \frac{1}{2} \Bigr)
\bigl(
\dot{\phi}^* (\tau) \dot{\phi} (\tau)
+ m^2 \phi^* (\tau) \phi (\tau) \bigr)
\end{equation}
is the quantum expectation value of the Hamiltonian
operator. It should be remarked that at the lowest order
the quantum effects enter the semiclassical Einstein equation
only through the quantum potential and there is no contribution
from the effective gauge potential.
The contribution from the quantum potential is still of the order
${\cal O} \Bigl(\frac{\hbar}{m_P} \Bigr)$, so we recover the
asymptotic semiclassical Einstein equation in previous papers \cite{kim1}.

The first order correction can be found from the equations:
\begin{equation}
{\bf \Omega}_{nn} c_n^{(0)} =  \frac{1}{\dot{a}^2}
\Biggl[\sum_{m \neq n} {\bf B}_{nm} {\bf B}_{mn} \Biggr] c_n^{(0)},
\end{equation}
and
\begin{equation}
{\bf \Omega}_{nk} c_n^{(0)} = - \frac{1}{\dot{a}^2}
\Biggl[ 2 {\bf B}_{nk} {\bf B}_{nn}
- \bigl( {\bf B}^2 \bigr)_{nk}
- i \Bigl( \dot{\bf B}_{nk} + \frac{\ddot{a}}{\dot{a}}
{\bf B}_{nk} \Bigr) \Biggr] c_n^{(0)}.
\end{equation}
Integrating the first order matter field equation
\begin{equation}
i \frac{\partial}{\partial \tau} c_{k}^{(1)}
= - \frac{2 \pi \hbar}{3 a}
 {\bf \Omega}_{kn} c_n^{(0)},
\end{equation}
we get
\begin{equation}
{\bf R}_{nn}^{(1)} =
\frac{\dot{c}_{n}^{(0)} + \frac{\hbar}{m_P}
\dot{c}_{n}^{(1)} }{
c_{n}^{(0)} + \frac{\hbar}{m_P}
c_{n}^{(1)}}
- \sum_{k} {\bf B}_{nk} \frac{
c_{k}^{(0)} + \frac{\hbar}{m_P}
c_{k}^{(1)} }{ c_{n}^{(0)} + \frac{\hbar}{m_P}
c_{n}^{(1)} }.
\label{f r}
\end{equation}
By substituting Eq. (\ref{f r}) into Eq. (\ref{frw 1}),
we obtain the semiclassical Einstein equation
up to the first order in $\frac{\hbar}{m_P}$.

\section{Conclusion}

The unitarity of quantum field in a curved spacetime
is an important issue in quantum gravity. In this paper, by
applying both the Born-Oppenheimer idea and the de Broglie-Bohm
interpretation, we derived in a different way the semiclassical
quantum gravity which consists of the nonlinear
gravitational equation and the time-dependent
functional Schr\"{o}dinger equation that
{\it does} preserve the unitarity of quantum field
without assuming any limit. In particular, we applied the de
Broglie-Bohm interpretation to the nonlinear gravitational equation,
whose real part is nothing but the Hamiltonian-Jacobi equation with
the quantum potential and the contributions from the
effective gauge potential of matter fields.
In an oscillating regime of gravity,
we were able to develop a perturbation method for the
semiclassical Einstein equation, which
include the quantum corrections of the  matter field.
Finally, we applied the perturbation method
to the quantum FRW cosmological model with a minimal
scalar field (inflaton) and obtained the semiclassical
Einstein equation (\ref{frw 1}). The right hand side of the
equation, the effective energy density,
is the sum of the expectation value of the Hamiltonian operator,
the higher order quantum corrections from the matter field and
the quantum potential coupled to the effective gauge potential.
Since the quantum corrections have the order of ${\cal O}
\Bigl(\frac{\hbar}{m_P} \Bigr)$, the semiclassical Einstein
equation reduces to the same asymptotic semiclassical Einstein equation
obtained already in Ref. \cite{kim1}.

This paper does not, however, resolve completely the problem of unitarity
of quantum field in quantum gravity and leave it to a more fundamental
theory. As pointed out earlier, bearing the de Broglie-Bohm interpretation
in mind, we made use of the freedom in the separation of the
Wheeler-DeWitt equation into the gravitational field equation
and the time-dependent Schr\"{o}dinger equation, and
derived explicitly the semiclassical quantum gravity that
does indeed {\it preserve} the unitarity of quantum evolution
of matter field at the expense of the nonlinearity
of the gravitational field equation.

\acknowledgments

The author would like to thank C. Kiefer
for suggesting developing the higher order quantum
corrections beyond the asymptotic limit
that motivated this paper and J.-Y. Ji
for explaining the de Broglie-Bohm interpretation of wavefunctions.
He also would like to thank
Professor K. S. Soh for continuous encouragement.
This work was supported in parts by
KOSEF under Grant No. 951-0207-056-2
and by the BSRI Program under Project No. BSRI-96-2427.

\end{document}